\documentclass {elsart}

\usepackage[tbtags]{amstex}
\usepackage{amscd}
\usepackage{epsfig}

\allowdisplaybreaks

\begin{document}

\date{11. March 1998}
\journal{Physica D}

\begin{frontmatter}

\title{Aftershocks in Coherent-Noise Models}

\author{C. Wilke},
\author{S. Altmeyer}, 
\author{T. Martinetz}
\address{Institut f\"ur Neuroinformatik, Ruhr-Universit\"at Bochum,\\
  D-44780 Bochum, Germany\\
e-mail: {\upshape
 \{{\ttfamily Claus.Wilke, Stephan.Altmeyer,
  Thomas.Martinetz}\}\\\ttfamily
$@$neuroinformatik.ruhr-uni-bochum.de}
}

\begin{abstract}
The decay pattern of aftershocks in the so-called 'coherent-noise' models
[M.~E.~J. Newman and K.~Sneppen, Phys. Rev.~E{\bf54}, 6226 (1996)] is
studied in detail. Analytical and numerical results show that the
probability to find a large event at time $t$ after an initial major
event decreases as $t^{-\tau}$ for small $t$, with the exponent $\tau$
ranging from 0 to values well above 1. This is in
contrast to Sneppen und Newman, who stated that the exponent is
about~1, independent of the microscopic details of the
simulation. Numerical 
simulations of an extended model [C.~Wilke, T.~Martinetz,
Phys. Rev.~E{\bf56}, 7128 (1997)]
show that the power-law is only a generic feature of the original
dynamics and does not necessarily appear in a more general
context.  Finally, the implications of the results to the modeling of
earthquakes are discussed. 

\end{abstract}

\end{frontmatter}

\section{Introduction}
Dynamical systems which display scale-free behaviour
 have attracted much interest in
recent years. Equilibrium thermodynamic systems do only exhibit
scale-free behaviour for a subset of the parameter space of measure
zero (the critical values of the parameters). Nevertheless, in nature
scale-free systems can be found in abundant variety
(earthquakes~\cite{Ma90}, avalanches in
rice-piles~\cite{Frette96}, infected people in
epidemics~\cite{Rhodes96}, jams in Internet traffic~\cite{Takayasu96},
extinction events in Biology~\cite{Sole96}, 
life-times of species~\cite{Keitt96} and many more. See
also~\cite{Dutta81} and the references therein). The  
origin of this abundance lies probably in the broad variety of
systems far from equilibrium that can be found in nature. With the onset of
non-equilibrium dynamics, new mechanisms come into play which seem to make
scale-free behaviour a generic feature of many systems. However,
unlike equilibrium thermodynamics, where scaling is 
thoroughly understood~\cite{Fischer74,Wilson75}, for non-equilibrium
dynamical systems  
there does not yet exist a unified theory of scale-free
phenomena (apart from non-equilibrium phase transitions). There do,
however, exist several distinct classes of 
systems with generic scale-free dynamic.

One of the first ideas to explain scale-free behaviour in a large
class of dynamical systems was the notion of Self-Organized Criticality (SOC)
proposed by Bak, Tang and Wiesenfeld in 1987~\cite{Bak87,Bak88}. They
proposed that 
certain systems with local interactions can, under the influence of a
small, local driving force, self-organize into a state with diverging
correlation length and therefore scale-free behaviour. This state
is similar to the ordinary critical state that arises at the critical
point in phase transitions, although no fine-tuning of parameters is
necessary to reach it. Since 1987 literally thousands of research
papers have been written concerning SOC (for a review
see~\cite{Bak94}), and many different dynamical 
systems have been called SOC (e.g. 
\cite{Bak93,Clar96,Drossel92,Olami92,Paczuski96}),
mostly just because they showed some 
power-law distributed activity patterns. Recently~\cite{Vespignani96}
it has become clear that several SOC models (sandpile models,
forest-fire models) can be understood in terms of ordinary
nonequilibrium critical phenomena (like
e.g.~\cite{Grassberger79}). Driving rate and dissipation act as
critical parameters. The critical value, however, is 0. Therefore, it
suffices to choose a small driving rate and dissipation to fine-tune
the system to the critical point, and this choice is usually implicit
in the definition of the model.

Scale-free behaviour does not, however, depend crucially on some sort
of critical phenomenon. A simple multiplicative stochastic process (MSP) of
the form
\begin{equation}\label{eq:multstoch}
  x(t+1)=a(t)x(t)+b(t)\,,
\end{equation}
where $a(t)$ and $b(t)$ are random variables, can produce a random
variable $x(t)$ with a probability-density function (pdf) with 
power-law tail~\cite{Levy96,Solomon96,Sornette97a,Sornette97b,Sornette97c}. In 
processes of this type, the power-law appears under 
relatively weak conditions on the pdf's of $a(t)$ and $b(t)$, thus
making the intermittend behaviour a generic feature of such
models. Applications of Eq.~(\ref{eq:multstoch}) can be found in
population dynamics with external sources~\cite{Sornette97c},
epidemics~\cite{Rhodes96}, price volatility
in economics~\cite{Mantegna95}, and others.

Another class of models with a very simple and robust
mechanism to produce scale-free behaviour has been introduced recently by
Newman and Sneppen~\cite{Newman96a}. These so called 'coherent-noise'
models consist of 
a large array of 'agents' which are forced to reorganize under
externally imposed stress. In their simplest form, coherent-noise
models do not have any interactions between the agents they consist 
of, and hence, certainly do not display criticallity. Nevertheless,
these models show a 
power-law distribution of the reorganization events with a wide range
of different exponents~\cite{Sneppen97}, depending on the special
implementation 
of the basic mechanism. Moreover they display power-law distributions
in several other quantities, e.g., the life-time distribution of the
agents. These models have been used to study 
earthquakes~\cite{Newman96a}, rice piles~\cite{Newman96a}, and biological
extinction~\cite{Newman96b,Newman97a,Newman97b,Wilke97}.  

Coherent-noise models have a feature that usually is not present in
SOC models and is never present in MSP's, which is the existence of
aftershocks. In 
most coherent-noise models the probability for a big event to occur is
very much increased immediately after a previous big event and then
decays with a power-law. This leads to a fractal pattern of
events that are followed by smaller ones which themselves are followed
by even smaller ones and so on.  In most SOC models and all MSP's, on
the contrary, the  
state of the system is statistically identical before and after a big
event. Therefore in these models no aftershocks are visible. 

The existence or non-existence of aftershock events should be easily
testable in natural systems. This could provide a means to decide what
mechanism is most likely to be the cause for scale-free behaviour in
different situations~\cite{Sneppen97}. But to achieve this it is
important to have a deep understanding of the decay-pattern of the aftershock 
events. 

The goal of the present paper is to investigate in detail the
aftershock dynamics of coherent-noise models. We concentrate
mainly on the original model introduced by Newman and Sneppen because
there can be obtained several analytical results. We find a power-law
decrease in time of the aftershocks' probability to
appear, as has been found 
already in~\cite{Sneppen97}. But unlike stated there, we can
show that 
the exponent does indeed depend on the microscopic details of the
simulation. We find a wide range of exponents, from~0 to values well
above~1, whereas in~\cite{Sneppen97} the authors report only
the value~1.

\section{The model}

We will now describe the model introduced by Newman and
Sneppen~\cite{Newman96a}.
The system consists of $N$ units or 'agents'. Every agent $i$ has a
threshold $x_i$ again external stress. The thresholds are initially chosen at
random from some probability distribution $p_{\rm thresh}(x)$. The
dynamics of the system is as follows:
\begin{enumerate}
  \item A stress $\eta$ is drawn from some distribution $p_{\rm
      stress}(\eta)$. All agents with $x_i\leq \eta$ are given new
      random thresholds, again from the distribution $p_{\rm
      thresh}(x)$.
  \item A small fraction $f$ of the agents is selected at random and
    also given new thresholds.
  \item The next time-step begins with (i).
\end{enumerate}
Step (ii) is necessary to prevent the model from grinding to a
halt. Without this random reorganization the thresholds of the agents
would after some time be well above the mean of the stress
distribution  and average stress could not hit any agents anymore.  

The most common choices for the threshold and stress distributions are
a uniform threshold distribution and some stress distribution that is
falling off quickly, like the exponential or the gaussian
distribution. Under these conditions (with reasonably small $f$) it is
guaranteed that the distribution of reorganization events that arises
through the dynamics of the system will be a power-law.

There are several possibilities to extend the model to make it more
general, without loss of the basic features. Two extensions that have
been studied by Sneppen and Newman~\cite{Sneppen97} are 
\begin{itemize}
 \item a lattice version where the agents are put on a lattice and
   with every agent hit by stress its nearest neighbours undergo
   reorganization, even if their threshold is above the current stress
   level.
 \item a 'multi-trait' version where, instead of a single stress,
   there are $M$ different types of stress, i.e.\ the stress becomes a
   $M$-dimensional vector $\boldsymbol{\eta}$. Accordingly, every 
   agent has a vector of thresholds $\mathbf{x_i}$.  An agent has to
   move in this model whenever at least 
   one of the components of the threshold vector is exceeded by the
   corresponding component of the stress vector.
\end{itemize}
An extension that is especially important for the application of
coherent noise models to biological evolution and the dynamics of mass
extinctions has been studied recently by Wilke and Martinetz~\cite{Wilke97}. In
biology it is not a good assumption to keep the number of agents (in this
case species) constant in time. Rather, species which go extinct
should be removed, and there should be a steady regrowth of new
species. In a generalized model, the system size is allowed to vary.
Agents that are hit by stress are removed from the system 
completely, but at the end of every time-step a number $\Delta N$ of
new agents is introduced into the system. Here $\Delta N$ is a
function of the actual system size $N$, the maximal system size
$N_{\max}$ and some growth rate $g$. Wilke and Martinetz have studied in
detail the function
\begin{equation}\label{eq:logistic_growth}
  \Delta N=\frac{N N_{\max} e^g}{N_{\max}+N(e^g-1)}-N\,,
\end{equation}
which resembles logistic growth. In the limit $g\rightarrow \infty$
their model reduces to the original one by Newman and Sneppen. In the
following we will refer to the original model as the
'infinite-growth version' and to the model introduced by Wilke and
Martinetz as the 'finite-growth version'.

\section{Analysis of the aftershock structure}

We base our analysis of the aftershock structure on the
meassurement-procedure proposed by Sneppen and
Newman~\cite{Sneppen97}. They drew a histogram of all the times whenever an
event of size~$\geq$ some constant $s_1$ happened after an initial event of
size~$\geq$ some constant $s_0$, for all events $\geq s_0$. Consequently,
we measure the frequency of events larger than $s_1$ occuring exactly
$t$ time-steps after an initial event larger than $s_0$, for all times
$t$. This means that we consider sequences of 
events in the aftermath of initial large events.
Normalized appropriately, our measurement gives just the probability to find an
event $\geq s_1$ at time $t$ after some arbitrarily chosen event $\geq
s_0$. For 
this to make sense in the context of aftershocks we usually have
$s_0>s_1$. Throughout the rest of this paper we use $s_0$ and $s_1$ as
percentage of the maximal system size. Therefore a value $s_0=1$ for
example means that we are looking for initial events which span the
whole system. 

We will denote the probability to find an
event of size $s\geq s_1$ at time $t$ after a previous large event by
$P_t(s\geq s_1)$. In order to keep the notation simple we omit the
constant $s_0$. It will be clear from the context what $s_0$ we use in
different situations. Note that $P_t(s\geq s_1)$ is not a probability
distribution, but a function of time $t$. Therefore, every increase or
decrease of $P_t(s\geq s_1)$ in time will indicate correlations
between the initial event and the subsequent aftershocks. For
$t\rightarrow\infty$ we expect all correlations to disappear, and
hence $P_t(s\geq s_1)$ to tend towards a constant.

It is possible to obtain some analytical results about the
probability $P_t(s\geq s_1)$ if we restrict ourself to the model with
infinite growth and a special choice for the threshold and stress
distributions. If 
not indicated otherwise, throughout the rest of this section we assume
$p_{\rm thresh}(x)$ to be uniform 
between 0 and 1, and the stress distribution to be exponential:
$p_{\rm stress}(\eta)=\exp(-\eta/\sigma)/\sigma$. 

 Furthermore, we focus on the case
$s_0=1$. That means that we are looking at the events in the aftermath
of an initial event of 'infinite' size, an event that spans the whole
system. This is a reasonable situation because we use a uniform
threshold distribution. In this case there is a finite probability
to generate a stress $\eta$ which exceedes the largest threshold, thus
causing the whole system to reorganize. For some of the examples
presented in Section~\ref{sec:numerical_results} the probability to
find an infinite event is even 
higher than $10^{-5}$. This probability can be considered relatively
large in a system where one has to do about $10^6-10^9$ time-steps to
get a good statistics. 

\subsection{Mean-field solution}

The exact way to calculate $P_t(s\geq s_1)$ is the
following. One has to compute the distribution
$\rho_{t-1}^{\eta_1,\eta_2,\dots,\eta_{t-1}}(x)$ which is the distribution
that arises if after the big event at time $t=0$ a sequence of stress
values $\eta_1,\eta_2,\dots,\eta_{t-1}$ is generated during the
following time steps. Then the equation
\begin{equation}\label{exact_cal_p}
  \int\limits_0^{x_t(\eta_1,\eta_2,\dots,\eta_{t-1},s_1)}\rho_{t-1}^{\eta_1,\eta_2,\dots,\eta_{t-1}}(x')\,dx'=s_1
\end{equation}
has to be solved. That gives the quantity
$x_t(\eta_1,\eta_2,\dots,\eta_{t-1},s_1)$, the threshold that has to
be exceeded by the stress at time $t$ to generate an event $\geq s_1$.
From the stress distribution one can then calculate the corresponding
probability $P_t^{\eta_1,\eta_2,\dots,\eta_{t-1}}(s\geq s_1)$. Finally
the average over all possible sequences
$\eta_1,\eta_2,\dots,\eta_{t-1}$ has to be taken to end up with
the exact solution for $P_t(s\geq s_1)$. Obviously there is no hope
doing this analytically. 

Instead of the exact solution for $P_t(s\geq s_1)$ we can calculate a
mean-field solution if we average over all possible sequences
$\eta_1,\eta_2,\dots,\eta_{t-1}$ before we solve
Eq. (\ref{exact_cal_p}). Note that in this context, the notion
mean-field does not stand for the average state of the system, which
does not tell us anything about aftershocks, but for the average
fluctuations typically found in a time-intervall $\Delta
t$. These average fluctuations are a measure for the return to the
average state, after a large event has caused a
significant departure from it.
Consequently, the mean-field solution is time-dependent. For $\Delta
t\rightarrow \infty$, the time-dependent mean-field threshold
distribution converges to the average threshold distribution, denoted
by $\bar\rho(x)$ in~\cite{Sneppen97}

In Appendix~\ref{A:master-equation}, we show that the averaging over
all fluctuations in a time intervall of $t$ time-steps equals to $t$
times iterating the master-equation. Therefore, to calculate the
mean-field solution for $P_t(s\geq s_1)$ we have to insert
$\rho_t(x)$, the $t$-th iterate of the master-equation, into
Eq.~(\ref{exact_cal_p}). The details of this calculation
are given in Appendix~\ref{A:master-iteration}.  

\subsection{Approximation for $\tau$}
\label{sect:approx_tau}

In this paragraph we will calculate the dependency of the exponent
$\tau$ on $s_1$ under the assumption that the probability to find
aftershocks decays indeed as a power-law, i.e.\ that we can assume
$P_t(s\geq s_1)\sim t^{-\tau}$.
A fairly simple argument shows that for the probability
$P_t(s\geq s_1)$ to 
decrease as a power-law the exponent $\tau$ must be proportional to
$1-s_1$ for $s_1$ not too small. Again we concentrate on
exponentially distributed stress only. 

We begin with an approximation of the quantity $x_t(s_1)$, which is
the average threshold at time $t$ above which a stress value must lie
to trigger an event of size $\geq s_1$. In the mean-field
approximation, $x_t(s_1)$ is defined by the equation
\begin{equation}
  \int\limits_0^{x_t(s_1)}\rho_t(x)\,dx = s_1\,.
\end{equation}
Because $\rho_t(x)$ and $s_1$ are normalized, we can rewrite this
equation (again we assume $p_{\rm
  thresh}(x)$ to be uniform between 0 and 1):
\begin{equation}\label{reversed_integral}
  \int\limits_{x_t(s_1)}^1\rho_t(x)\,dx = 1-s_1\,.
\end{equation}
For the most reasonable stress distributions the
distribution of the agents $\rho_t(x)$ forms a
plateau in the region close to $x=1$. Therefore for sufficient large
$s_1$ we can
approximate the integral in Eq. (\ref{reversed_integral}) by substituting
$\rho_t(x)$  with its value at $x=1$, which is $\rho_t(1)$.
Eq. (\ref{reversed_integral}) then becomes
\begin{equation}
  \Big(1-x_t(s_1)\Big)\rho_t(1)=1-s_1\,.
\end{equation}
The values $\rho_t(1)$ are a function of $t$. We define
\begin{equation}\label{Rtdef}
  R(t):=\rho_t(1)
\end{equation}
and find for $x_t(s_1)$:
\begin{equation}
  x_t(s_1)=\frac{s_1-1+R(t)}{R(t)}\,.
\end{equation}
The probability $P_t(s\geq s_1)$ now becomes
\begin{equation}\label{P_texp}
  P_t(s\geq s_1)=\exp\Big(-\frac{x_t(s_1)}{\sigma}\Big)=
         \exp\Big(- \frac{s_1-1+R(t)}{\sigma R(t)}\Big)\,.
\end{equation}

The principal idea to derive a relation between $\tau$ and $s_1$ is 
as follows. The function $R(t)$ is obviously independent of $\tau$ and
$s_1$. We make the ansatz $P_t(s\geq s_1)\sim t^{-\tau}$, rearrange
Eq. (\ref{P_texp}) for $R(t)$ and then get a condition on $\tau$ and
$s_1$ since they should cancel exactly. Hence we have to solve the
equation
\begin{equation}\label{P_texp2}
  at^{-\tau} = 
         \exp\Big(- \frac{s_1-1+R(t)}{\sigma R(t)}\Big)\,,
\end{equation}
where $a$ is the constant of proportionality. We take the logarithm on
both sides to get
\begin{equation}
  \ln a- \tau\ln t=- \frac{s_1-1+R(t)}{\sigma R(t)}
\end{equation}
and finally
\begin{equation}
  R(t)=\frac{1-s_1}{1+\sigma\ln a-\tau\sigma\ln t}\,.
\end{equation}
This is of the form $c_1/(c_2-\ln t)$, where 
\begin{equation}\label{c_1def}
 c_1=\frac{1-s_1}{\tau\sigma}
\end{equation}
and
\begin{equation}\label{constant2}
 c_2=\frac{1+\sigma\ln a}{\tau\sigma}\,.
\end{equation}
For every function of the form $c_1/(c_2-\ln t)$, the constants 
$c_1$ and $c_2$ are unique, as can be seen if we write
\begin{equation}
  \frac{c_1}{c_2-\ln t}=\frac{c_1}{\ln\frac{\exp c_2}{t}}\,.
\end{equation}
A change in $c_2$ results in a rescaling of the variable $t$, while a
change in $c_1$ results in a rescaling of the whole
function. Consequently, neither $c_1$ nor $c_2$ can depend on $\tau$
or $s_1$. This can be seen as follows. If, e.g., $c_1$ depended on
$s_1$, then a change in $s_1$ would rescale the function $R(t)$. But
this function is independent of $s_1$  according
to its definition (Eq.~(\ref{Rtdef})). Hence $c_1$ must be independent
of $s_1$ in itself. A similar argument holds for the variable $c_2$.
Therefore, $s_1$ and $\tau$ must cancel exactly in
Eq.~(\ref{c_1def}). This leads to the condition
\begin{equation}\label{tau_prop_s1}
  \tau=\frac{1-s_1}{\sigma c_1}\sim (1-s_1)\,.
\end{equation}

Up to now we have not done any assumptions about the size of the first
big event after which we are measuring the subsequent
aftershocks. Therefore the proportionality $\tau\sim(1-s_1)$ should
hold in general, as long as $s_1$ is not too small. If we additionally
assume the inital event to have infinite size ($s_0=1$) we can easily
calculate the constant $a$ in Eq. (\ref{P_texp2}). The meaning of this
constant is the probability to get an event of size $\geq s_1$
immediately after the initial big event, as can be seen by setting
$t=1$:
\begin{equation}
  P_1(s\geq s_1)=a1^{-\tau}=a\,.
\end{equation}
For the case $s_0=1$ the distribution of thresholds $\rho_0(x)$ is
uniform and thus
\begin{equation} \label{const_a}
  a=\exp(-\frac{x_1(s_1)}{\sigma})=\exp(-\frac{s_1}{\sigma})\,.
\end{equation}
With Eqs. (\ref{P_texp}), (\ref{P_texp2}), (\ref{c_1def}), and
(\ref{const_a}) we can write the probability $P_t(s\geq s_1)$ as
\begin{equation}\label{P_tfinal}
  P_t(s\geq s_1) = e^{-s_1/\sigma}t^{-(1-s_1)/(\sigma c_1)}\,.
\end{equation}
In Section \ref{sec:numerical_results} we will find numerically that
$c_1=\sigma^{-1}$, and therefore $\tau=1-s_1$.

\subsection {Limiting cases}
\label{sect:limiting_cases}

For two limiting cases we can deduce the behaviour of the exponent
$\tau$
regardless of the stress distribution. We begin with the
case $s_0=1$, $s_1\rightarrow 1$. From Eq.~(\ref{tau_prop_s1}) we find
that $\tau\rightarrow 0$ as $s_1\rightarrow 1$ under the assumption of
an exponential stress distribution. But this result is more
general. For $s_1=1$ the probability $P_t(s\geq s_1)$ reads simply
\begin{equation}
  P_t(s\geq 1)=\int\limits_1^\infty\!dx\, p_{\rm stress}(x)
\end{equation}
and hence is constant in time. Consequently we have $\tau=0$ for any
stress distribution. From continuity we have $\tau\rightarrow 0$ as
$s_1\rightarrow 1$. 

A similar argument holds when either $s_0$ or $s_1$ approaches 0. For
$s_0=0$, the probability $P_t(s\geq s_1)$ reduces to the
mean probability to find an event of size $s\geq s_1$. Hence $\tau=0$.
For $s_1=0$, the probability $P_t(s\geq s_1)$ becomes 1,
because an event of size at least zero can be found in every time
step. Hence also in this case $\tau=0$. From continuity we have again
$\tau\rightarrow 0$ as either $s_0\rightarrow 0$ or $s_1\rightarrow 0$.

\section{Numerical results}
\label{sec:numerical_results}

In the previous section we have focused on the behaviour of the
system in the aftermath of an infinite event. This situation is not 
only analytically tractable, but it also makes it simpler to obtain
good numerical results. If we want to measure the probability to find
aftershocks following events exceeding some finite but large $s_0$,
we have to wait a long time for every single measurement since the
number of those large events vanishes with a power-law. This makes it
hard to get a good statistics  within a reasonable amount of computing
time. If, on the other hand, we focus on the situation of an infinite
initial event, we can simply initialize the agents with the uniform
threshold distribution, take the measurement up to the time $t$ we
are interested in and repeat this procedure until the desired accuracy is
reached. Unless stated otherwise, the results reported below have been
obtained in this way, and with exponentially distributed stress.

\begin{figure}
\centerline{
\epsfig{file={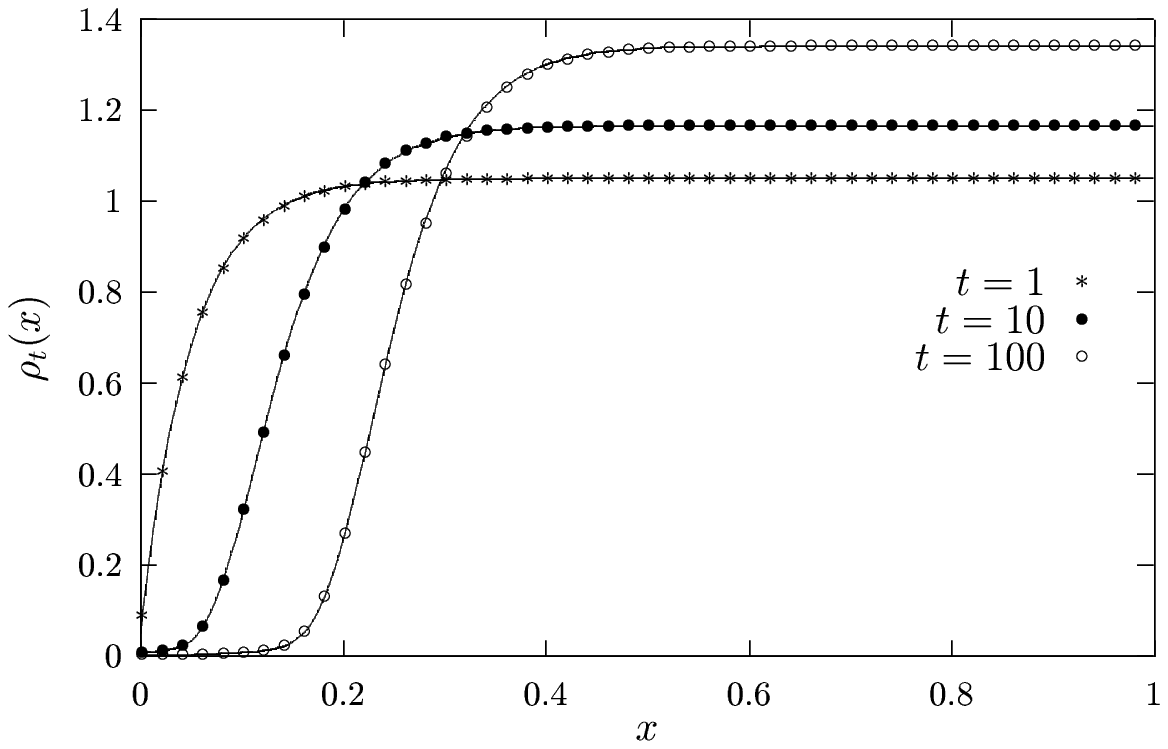}}
}
\caption{The average distribution of agents at time $t=1$, $t=10$, and
  $t=100$ after the initial event of size $\infty$. The solid line is
  the analytical result from the iteration of the master-equation, the
  dots show the simulation results. Parameters where $\sigma=0.05$ and
  $f=10^{-5}$}\label{Fig_rho_t}
\end{figure}

The $t$-th iteration of the master-equation should give exactly the average
distribution of the agent's thresholds at time $t$. In
Fig.~\ref{Fig_rho_t} it can be seen that this is indeed the case. The
points, which represent simulation results, lie exactly on the solid
lines, which stem from the exact analytical calculation.

\begin{figure}
\centerline{
\epsfig{file={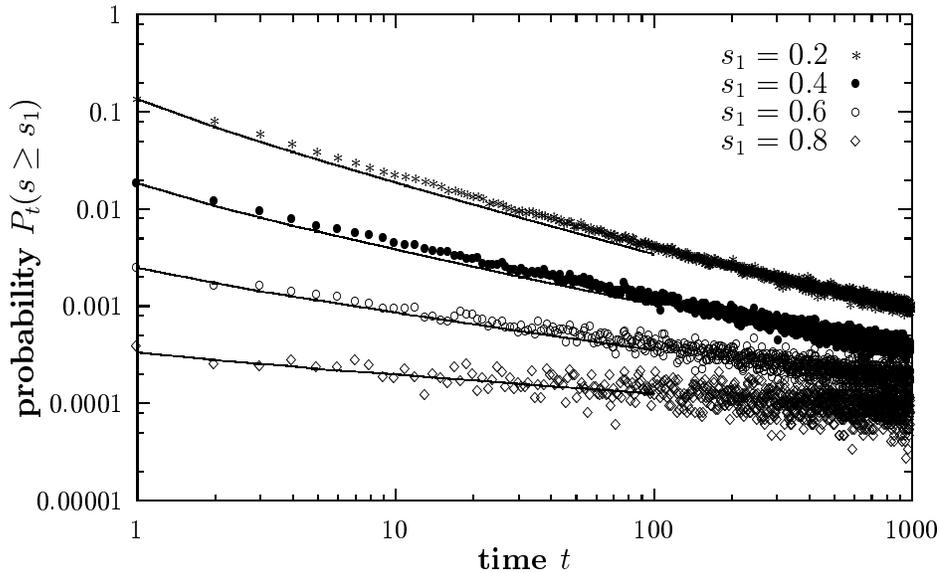}}
}
\caption{The probability $P_t(s\geq s_1)$ for
  $\sigma=0.1$ and $f=10^{-5}$. The solid lines 
  show the mean-field approximation, the dots show the simulation
  results.}\label{Fig_mf1}  
\end{figure}

The mean-field approximation for $P_t(s\geq s_1)$ should be valid if
the agent's distribution at time $t$ does not fluctuate too much about
the average distribution $\rho_t(x)$. Since there are many more small
events than big ones the fluctuations should occur primarily in the
region of small $x$. Consequently we expect the mean-field
approximation to be valid for large $s_1$, and to break down for small
$s_1$. Fig. \ref{Fig_mf1} shows that already for moderately large $s_1$ the
mean-field approximation captures the behaviour of $P_t(s\geq s_1)$,
with a slight tendency to underestimate the results of the
measurement. Note that the statistics is becoming worse with
increasing $s_1$ due to the rapidly decreasing probability to find any
events of size $\geq s_1$ for large $s_1$.

\begin{figure}
\centerline{
\epsfig{file={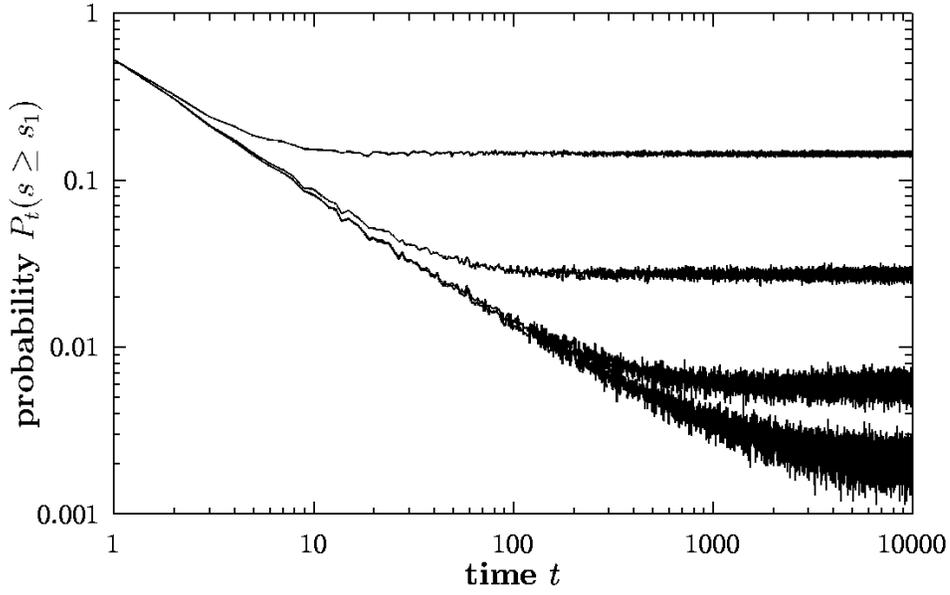},width=12.5cm}
}
\caption{The probability $P_t(s\geq s_1)$ in simulations with $\sigma=0.1$,
  $s_1=0.06$ and several different 
  values for $f$ (from bottom to top: $f=10^{-4}$, $f=10^{-3}$,
  $f=10^{-2}$, $f=10^{-1}$). }\label{Fig_limf}  
\end{figure}

In Fig. \ref{Fig_limf} a measurement of the probability $P_t(s\geq
s_1)$ is presented for a 
number of simulations with different values of the parameter $f$. 
As it can be seen, the parameter $f$ does not affect the exponent of
the power-law, but limits the region where scaling can be
observed. Note the difference between the effect seen here and typical
cut off effects in the theory of phase transitions. The quantity $P_t(s\geq
s_1)$ is not a probability distribution, but a time dependent
probability, which tends towards a constant for
$t\rightarrow\infty$. Therefore, we do not see an exponential decrease
at the cut off timescale. Instead, the probability $P_t(s\geq
s_1)$ levels out to the time-averaged value $P(s\geq s_1)$, which is
the average probability to find events of size $s\geq s_1$.

\begin{figure}
\centerline{
\epsfig{file={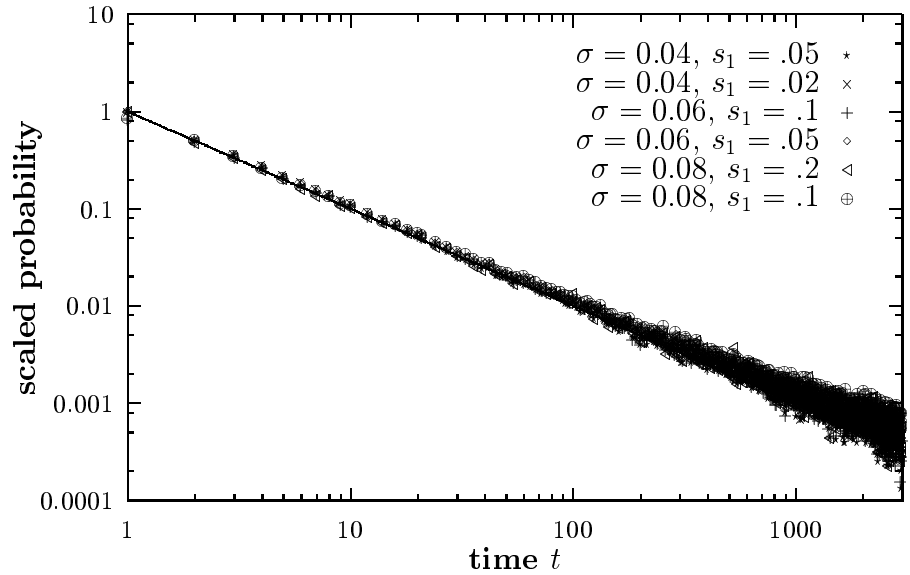}}
}
\caption{The rescaled probability $P_t(s\geq s_1)$. The solid line
  shows a $t^{-1}$ decrease for comparison.}\label{Fig_scaledP}
\end{figure}

In section \ref{sect:approx_tau} we showed that $\tau\sim
1-s_1$, under the condition of a sufficiently large $s_1$.
Simulations indicate that the constant of proportionality is just 1,
which means the constant $c_1$ in Eq.~(\ref{c_1def}) equals
$\sigma^{-1}$.  
If this hypothesis is true, Eq.~(\ref{P_tfinal}) becomes
\begin{equation}\label{Eq:conj1} 
  P_t(s\geq s_1) = e^{-s_1/\sigma}t^{-(1-s_1)}\,.
\end{equation}
This means, a rescaling of the form
\begin{equation}
  P_{\rm scaled} = \left(\frac{P_t(s\geq s_1)}{e^{-s_1/\sigma}}
      \right)^{1/(s_1-1)}
\end{equation}
should yield a functional dependency $P_{\rm scaled}(t)
=t^{-1}$. Fig. \ref{Fig_scaledP} shows the results of
such a rescaling for different $\sigma$ and $s_1$. All the data-points
lie exactly on top of each other in the region where the statistics is good
enough (about $t<100$). We find that for $\sigma$ up to 0.1,
Eq. (\ref{Eq:conj1}) is very accurate for $s_1$ between about 0.1 and
1. With smaller $\sigma$, Eq. (\ref{Eq:conj1}) holds even for much
smaller $s_1$. 

\begin{figure}
\centerline{
\epsfig{file={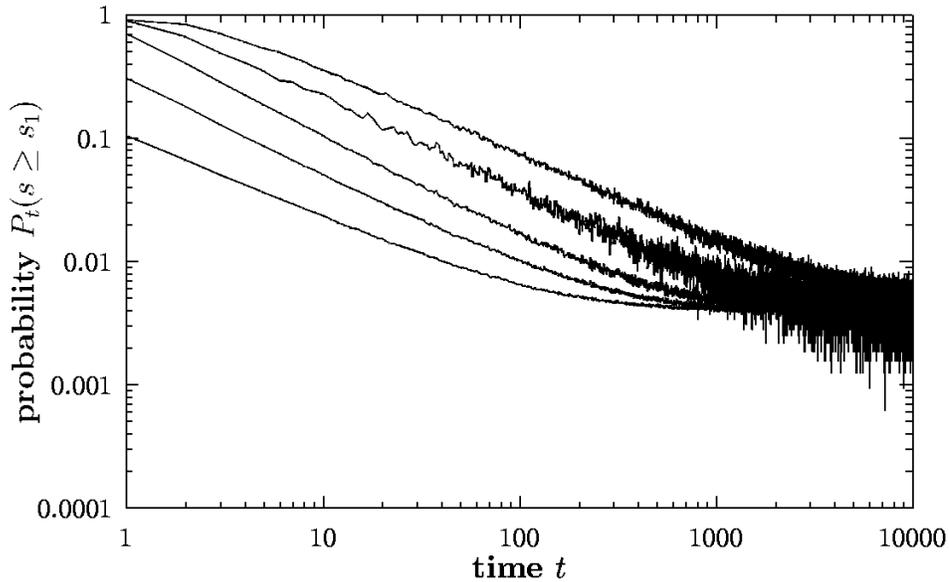}, width=12.5cm}
}
\caption{The probability $P_t(s\geq s_1)$ after an initial finite
  event. Parameters are 
  $\sigma=0.05$, $f=10^{-5}$, $s_1=3\times10^{-4}$, and, from bottom
  to top, $s_0=5\times10^{-4}$, 
  $s_0=2\times10^{-3}$, $s_0=0.01$, $s_0=0.1$ and
  $s_0=1$.}\label{Fig_finites0}   
\end{figure}

The situation becomes more complicated if we study the sequence of aftershocks
caused by an initial event of finite size. The probability to find an
event of size $s\geq s_1$ after some initial event of size $s\geq s_0$
decreases with a power-law, but the exponent is not a simple 
function of $s_1$. Rather, it depends on $s_0$ as well. In
Fig.~\ref{Fig_finites0} we have displayed the results of a
measurement with $s_1=3\times10^{-4}$ and several different $s_0$,
ranging from $5\times10^{-4}$ to 1. The curve for $s_0=1$ has been
obtained with the method described at the beginning of this
section. We find that the aftershocks' decay pattern for $s_0<1$
continuously approaches the one for $s_0=1$ as $s_0\rightarrow
1$. This shows that it is indeed justified to study the system in the
aftermath of an infinite initial event and then to extrapolate to
finite but large initial events. Note that in Fig.~\ref{Fig_finites0},
$s_1$ is so small that Eq.~(\ref{Eq:conj1}) does not hold anymore.

\begin{figure}
\centerline{
\epsfig{file={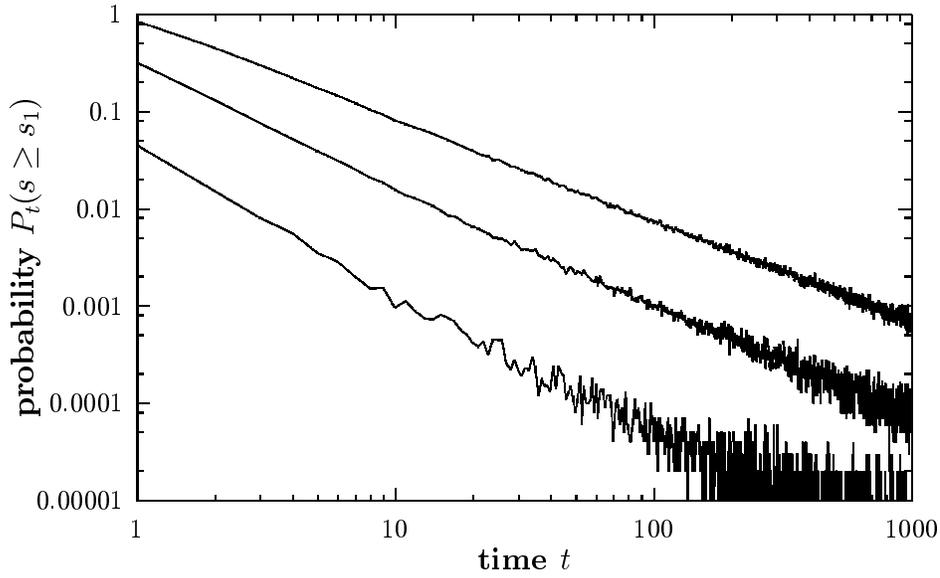}}
}
\caption{The probability $P_t(s\geq s_1)$ after an initial infinite
  event in a simulation with gaussian stress. Parameters are 
  $\sigma=0.1$, $f=10^{-5}$, and, from bottom to top, $s_1=0.2$,
  $s_1=0.1$, $s_1=0.02$. In a simulation with gaussian stress the
  distribution is getting steeper with increasing $s_1$.}\label{Fig_gauss.1}  
\end{figure}

Sneppen and Newman have argued that the decay pattern of the aftershocks is
independent of the respective stress distribution. Our results do not
support that. Despite the fact that the exponent of the power-law
seems to be independent of $\sigma$ in the case of exponential
stress, as we could show above, the exponent is not independent of the
functional 
dependency of the stress distribution. If we impose, for example,
gaussian stress with mean 0 and variance $\sigma$, we find
(Fig.~\ref{Fig_gauss.1}) exponents larger than 1 for moderate
$s_1$. We do
event find a qualitatively new behaviour of the system.
The exponent is getting larger with
increasing $s_1$, as opposed to the exponent getting smaller for
exponential stress. Of course, this can only be true for intermediate
$s_1$. Ultimately, we must have $\tau\rightarrow 0$ for
$s_1\rightarrow 1$.

\begin{figure}
\centerline{
\epsfig{file={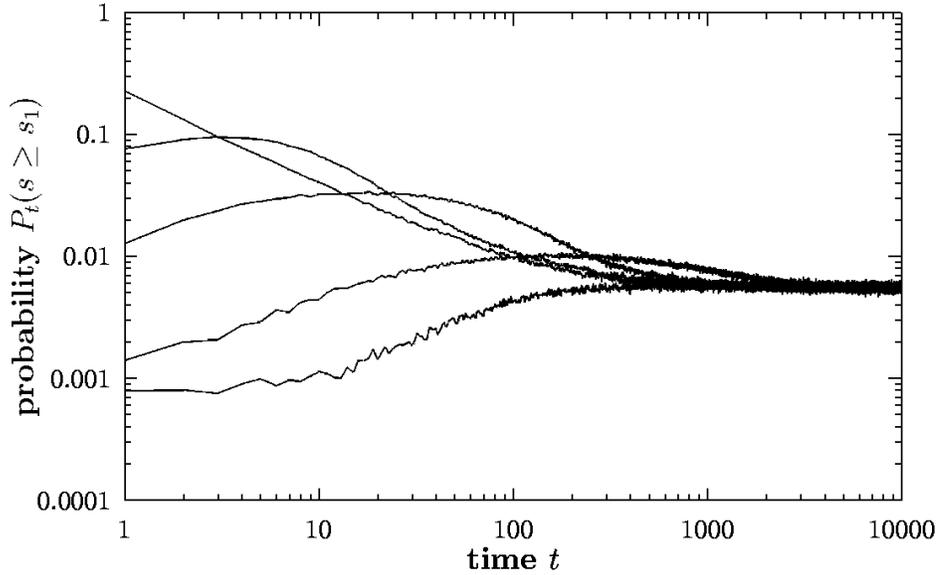}, width=12.5cm}
}
\caption{The probability $P_t(s_1\geq 3\times10^{-4})$ after an event of
  size $s_0\geq 0.001$ 
  in a simulation with finite logistic growth. Parameters are 
  $\sigma=0.5$, $f=10^{-5}$, and, from bottom to top,
  $g=2\times10^{-4}$, $g=2\times10^{-3}$, $g=2\times10^{-2}$
  $g=2\times10^{-1}$, $g=10$. 
  A power-law can only be seen for relatively large growth rates. For
  small growth-rates, the probability to find aftershocks is reduced
  significantly.}\label{Fig_grow}   
\end{figure}

Finally, we present some results about systems with finite growth. In
these systems, there exists some competitive dynamics between the
removal of agents with small thresholds through stress and their
regrowth. Aftershocks appear in the infinite growth model because the
reorganization event leaves a larger 
proportion of agents in the region of small thresholds, thus 
increasing the probability for succeding large events. In the finite
growth version, on the contrary, this can happen only if the regrowth
is faster than the constant removal of agents with small thresholds
through average stress. If the regrowth is too slow,
the probability to find large events actually is decreased in the
aftermath of an initial large event. The interplay between these two
mechanisms is shown in Fig.~\ref{Fig_grow}. The regrowth of
species is done according to Eq.~(\ref{eq:logistic_growth}). For a small
growth-rate $g$, the probability to find aftershocks is reduced
significantly, and it 
aproaches the equilibrium value after about 100 time-steps. With
larger $g$, the probability $P_t(s\geq s_1)$ increases in time until
a maximum well above the 
equilibrium level is reached, and then decreases again. 
The maximum moves to the left to earlier times $t$ with increasing
$g$. When $g$ is so large that the maximum coincides with the
post initial event, the original power-law is restored. Note that, as
in the case of infinite growth, the measurement depends on the choice
of the parameters $s_0$ and $s_1$. Consequently, with a different set
of parameters the curves will look different, and the maximum will
appear at a different time. Nevertheless, we find that the
qualitative behaviour is not altered if we change $s_0$ or $s_1$.

\begin{figure}
\centerline{
\epsfig{file={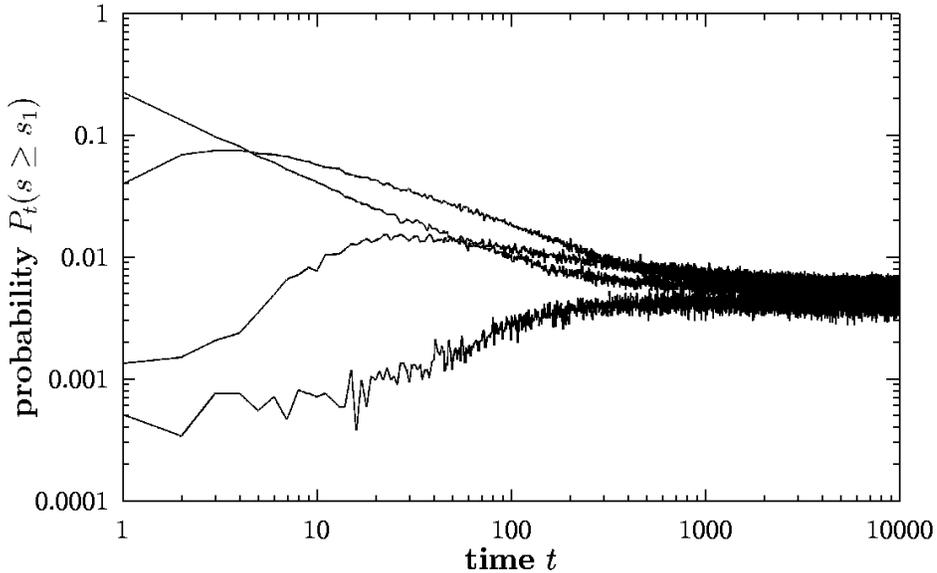}, width=12.5cm}
}
\caption{The probability $P_t(s_1\geq 3\times10^{-4})$ after an event of
  size $s_0\geq 0.001$ 
  in a simulation with finite linear growth. Parameters are 
  $\sigma=0.5$, $f=10^{-5}$, and, from bottom to top,
  $g=10^{-5}$, $g=\times10^{-4}$, $g=\times10^{-3}$, $g=10$. 
  Aftershocks are seen for much smaller growth rates than in the
  version with logistic growth.}\label{Fig_growl}   
\end{figure}

Instead of logistic growth, we can also think of linear growth, i.e.\
$\Delta N=gN_{\max}$,
where $g$ is again the growth rate. In order to keep the system
finite, we stop the regrowth whenever the system size $N$
exceedes the maximal system size $N_{\max}$. In such a system,
aftershocks can be seen for much smaller growth rates
(Fig.~\ref{Fig_growl}). Note that apart from the growth rate, all
other settings are identical in Fig.~\ref{Fig_grow} and
Fig.~\ref{Fig_growl}. Linear growth refills the system much quicker
than logistic growth with the same growth rate. Therefore the time
intervall in which aftershocks are suppressed is much shorter for
linear growth.

\section{Conclusion}

We could show in the present paper that the decay pattern of the  aftershock
events depends on the details of
the measurement. Although the qualitative features remain the same for
different parameters $s_0$ and $s_1$ (e.g.\ a power-law decrease in
the infinite-growth version), the quantitative features vary to a
large extend.  The exponent $\tau$ of the power-law is
significantly affected by an alteration of $s_0$ or $s_1$. Therefore
the measurement-procedure proposed by Sneppen and Newman can reveal
the complex structure of aftershock-events only if the
change of the measured decay pattern with varying $s_0$ and $s_1$ is
recorded over a reasonably large intervall of different values. This
should be considered in a possible comparison between the
aftershocks' decay pattern from a model and from experimental data. A
more in-depth analysis could probably be achieved with the formalism
of multifractality (see e.g.~\cite{Pastor-Satorras97}).

We found the aftershocks' decay pattern to vary with different stress
distributions. This is in clear contrast to Sneppen and Newman. They
reported a power-law with exponent
$t\approx 1$ for the infinite growth version, independent of the
respective stress distribution they 
used. The question remains why Sneppen and Newman measured such an
exponent in all their simulations. The answer to this lies in the fact
that they did only simulations with $s_0<1$. For the reasons explained
at the beginning of Section~\ref{sec:numerical_results}, under this
condition one has to choose a relatively small $s_0$, and accordingly,
a very small $s_1$. This causes the
measurement almost inevitably to lie in the intermediate region
between the limiting cases of Sec.~\ref{sect:limiting_cases}. In this
region, for the 
most reasonable stress distributions and a large array of different
values for $s_0$ and $s_1$, we find indeed exponents
around~1.

The application of coherent-noise models to earthquakes has been
discussed in~\cite{Newman96a}. Two very important observations
regarding earthquakes, the Gutenberg-Richter
law~\cite{Gutenberg54,Gutenberg56} and Omori's
law~\cite{Omori1894a}, can
be explained easily with a coherent-noise model. The Gutenberg-Richter
law states that earthquake magnitudes are distributed according to a
power-law. Omori's law, which interests us here, is a similar
statement for the aftershocks' decay pattern of earthquakes. In the data
from real earthquakes, the number of events larger than a certain
magnitude $M_1$ decreases as $t^{-\tau}$ in the aftermath of a large
initial earthquake. Consequently, we can only apply infinite-growth
coherent-noise models to earthquakes. But this is certainly no
drawback, since we expect the thresholds against movement at various
points along a fault (with which we identify the agents of the
coherent-noise model) to reorganize almost instantaneously during an
earthquake. 

The exponent $\tau$ is not universal, but can vary significantly, e.g.,
in~\cite{Ma90} from $\tau=0.80$ to $\tau=1.58$. This would
cause problems if the statement was true that for coherent-noise
models we have $\tau\approx 1$. But as we could show above, the
exponent can assume values in the observed range,
depending on the stress distribution, the size of the initial event,
and the lower cut-off (which we called $M_1$ for earthquakes and $s_1$
throughout the rest of the paper). 

For a further comparison, it should be interesting to study the
dependency of the exponent $\tau$ on variation of the cut-off $M_1$ in
real data. We are only aware of a single work where that has been
done~\cite{Hastings93}. Interestingly, the authors do not find a clear
dependency $\tau(M_1)$. Nevertheless, this is not a strong evidence
against coherent-noise models, since the aftershock series analysed
in~\cite{Hastings93} consists mainly of very large earthquakes with magnitude
between 6 and about 8, which does not allow a sufficient variation of
$M_1$. Statistical variations in the exponent $\tau$ are probably
larger for this aftershock series than the possible variations because
of an assumed $\tau(M_1)$ dependency. 

Numerical simulations of the finite-growth version have revealed a
much more complex structure of aftershock events than present in the
infinite-growth version. The competition between regrowth of agents
and agent removal through reorganization events leads to a pattern
where the probability to find events after an initial large event is
suppressed for short times, enhanced for intermediate times and then
falls off to the background level for long times. This observation
can be important for the application of coherent-noise models to
biological extinction. It might be possible to identify a time of
reduced and a time of enhanced extinction activity in the aftermath of
a mass extinction event in the fossil record. This would be a good
indication for biological extinction to be dominated by external
influences (coherent-noise point of view) rather than by coevolution (SOC
point of view). 

\begin{ack}
We thank Mark Newman for interesting conversations about
coherent-noise models.
\end{ack}

\begin{appendix}

\section{Rederivation of the master-equation}
\label{A:master-equation}

In this appendix we are interested in the average state a
coherent-noise system will be found in several time-steps after some
initial state with threshold distribution $\rho_{t_0}(x)$. Our
calculations will lead to a rederivation of the master-equation for
coherent-noise systems.
Although a master-equation has
already been given for the infinite-growth version and has been
generalized to the finite-growth version, these master-equations have
not 
been derived in a stringent way, but just have been written down
intuitively. Our calculation will confirm the main terms of the
previously used equations, but we will find an additional correcting
term that becomes important for large $f$. 

Consider the case of infinite growth. At time $t_0$ the
threshold-distribution may be $\rho_{t_0}(x)$. We  
construct the distribution $\rho_{t_0+1}^\eta(x)$, which is the
distribution that arises at time $t_0+1$ if a stress $\eta$ is
generated at time $t_0$. A stress $\eta$ will cause a proportion
$s_\eta=\int_0^\eta\!dx\,\rho_{t_0}(x)$ of the agents to move. We have to
distinguish two regions. For $x<\eta$, all agents are removed. Then they
are redistributed according to $s_\eta p_{\rm thresh}(x)$. A small
fraction $f$ of the agents is then mutated, which results in
\begin{equation}\label{rhostep}
  \rho_{t_0+1}^\eta(x)=(1-f)s_\eta p_{\rm thresh}(x)+fp_{\rm thresh}(x)
    \quad;\; x<\eta\,.
\end{equation}
For $x\geq\eta$, the redistribution of the agents gives
$\rho_{t_0}(x)+s_\eta p_{\rm thresh}(x)$. With the subsequent mutation
we obtain: 
\begin{equation}
   \rho_{t_0+1}^\eta(x)=(1-f)\Big(\rho_{t_0}(x)+s_\eta p_{\rm
    thresh}(x)\Big)+fp_{\rm thresh}(x) 
    \quad;\; x\geq\eta\,.
\end{equation}

We take the average over $\eta$ to get the distribution
$\rho_{t_0+1}(x)$ that will on average be found one time-step after
$\rho_{t_0}(x)$:
\begin{align} \label{master-begin}
  \rho_{t_0+1}(x)&=\int\limits_0^\infty \!d\eta\, p_{\rm
  stress}(\eta)\rho_{t_0+1}^\eta(x)\notag\\
 &= \int\limits_0^\infty \!d\eta\, p_{\rm stress}(\eta) p_{\rm thresh}(x)
        \Big[(1-f)\int\limits_0^\eta 
       \rho_{t_0}(x')dx'+f\Big] \notag\\
 &\qquad + \int\limits_0^x \!d\eta\, p_{\rm stress}
  (\eta) \rho_{t_0} (x)(1-f) \notag\\
 &=  p_{\rm thresh}(x)\Big[f+(1-f)\int\limits_0^\infty \!d\eta\,
  p_{\rm stress}(\eta) \int\limits_0^\eta 
       \rho_{t_0}(x')dx'\Big] \notag\\
 &\qquad + \rho_{t_0} (x)(1-f)(1-p_{\rm move}(x))\,.
\end{align}
Here, $p_{\rm move}(x)$ is the probability for an agent with threshold
$x$ to get hit by stress, viz. $p_{\rm
  move}(x)=\int_x^\infty\!dx'\,p_{\rm stress}(x')$. 
To proceed further we have to interchange the order of integration in
the remaining double integral. Note that $\int_0^\infty
\!d\eta\,\int_0^\eta \!dx' = \int_0^\infty
\!dx'\,\int_{x'}^\infty \!d\eta$, and therefore
\begin{align} \label{master_cont}
 \rho_{t_0+1}(x) &=  p_{\rm thresh}(x)\Big[f+(1-f)\int\limits_0^\infty \!dx'\,
  \int\limits_{x'}^\infty \!d\eta\, p_{\rm stress}(\eta)
       \rho_{t_0}(x')\Big] \notag\\
 &\qquad + \rho_{t_0} (x)(1-f)(1-p_{\rm move}(x)) \notag\\
 &=  p_{\rm thresh}(x)\Big[f+(1-f)\int\limits_0^\infty \!dx'\,
         \rho_{t_0}(x')p_{\rm move}(x')\Big] \notag\\
 &\qquad + \rho_{t_0} (x)(1-f)(1-p_{\rm move}(x)) \notag\\
 &=  p_{\rm thresh}(x)\int\limits_0^\infty \!dx'\,\Big(
        f+(1-f) p_{\rm move}(x')\Big)\rho_{t_0}(x') \notag\\
 &\qquad + \rho_{t_0} (x)(1-f)(1-p_{\rm move}(x)) 
\end{align} 
We are thus led to the master-equation
\begin{equation}\label{master-equation}
  \Delta\rho_t(x)=\Big(-f-p_{\rm move}(x)+fp_{\rm
  move}(x)\Big)\rho_t(x)+ Ap_{\rm thresh}(x)\,, 
\end{equation}
where $A$ is the normalization constant $\int_0^\infty \!dx'\,(
 f+(1-f) p_{\rm move}(x'))\rho_t(x')$.

We notice the appearance of the term $fp_{\rm move}(x)$ which was not
present in the master-equation used by Sneppen and Newman. The term
arises if one takes into account the fact that the agents which are
hit by  stress get new thresholds before the mutation
takes place. Therefore every agent with threshold $x$ has the
probability $fp_{\rm 
  move}(x)$ to get two new thresholds in one time-step. But obviously 
this is exactly the same as geting only one new threshold. Consequently, the
term   $fp_{\rm move}(x)$ has to be present to avoid double-counting
of those agents which are hit both by stress and by
mutation. Nevertheless, this term does not affect the scaling
behaviour of the system, because the derivation of the event size
distribution in~\cite{Sneppen97} has been done under the assumption
$f\ll 1$.

Eq. (\ref{master_cont}) gives the average state of the system one
time-step after the initial state $\rho_{t_0}(x)$. If we are
interested in the average state $t$ time-steps after the initial
state, we have to repeat the calculations in
Eqs. (\ref{rhostep})-(\ref{master_cont}) $t$ times. Since all
averages in these calculations can be taken independently, this is
exactly the same as $t$ times iterating the master-equation
(\ref{master-equation}). 

\section{Calculation of the mean-field solution.}
\label{A:master-iteration}

 We assume that at time $t=0$ a big event takes place
and produces the distribution $\rho_0(x)$. If we apply the
master-equation (\ref{master-equation}) $t$ times to this distribution
$\rho_0(x)$, we will end up with a distribution $\rho_t(x)$ that tells
us the average state of the system  at time $t$ after the
big event. 

In the following we will use
\begin{equation}
  T(x):=(1-f)(1-p_{\rm move}(x))
\end{equation}
and write $A_t$ for the normalization constant that appears on the
right-hand side of Eq.~(\ref{master-equation}) at time $t$.
The average distribution at time $t$ then becomes
\begin{align}\label{master-T}
  \rho_t(x)&=T(x)\rho_{t-1}(x)+A_t p_{\rm thresh}(x) \notag\\
   &= T^t(x)\rho_0(x) + \sum\limits_{k=1}^t T^{t-k}(x)A_k p_{\rm thresh}(x) \,.
\end{align}
We integrate both sides of Eq.\ (\ref{master-T}) and find a
recursion relation for the constants $A_t$:
\begin{equation}\label{At-recursion}
  A_t=1-\int\limits_0^\infty T^t(x)\rho_0(x)dx+\sum\limits_{k=1}^{t-1}A_k
       \int\limits_0^\infty T^{t-k}(x)p_{\rm thresh}(x)dx\,.
\end{equation}
All integrals can be calculated analytically for a special choice of
the threshold and stress distributions. As threshold distribution, we
choose the uniform distribution $p_{\rm thresh}(x)=1;\; 0\leq x<1$, and
as stress distribution we choose the exponential distribution $p_{\rm
  stress}(\eta)=\exp(-\eta/\sigma)/\sigma$. Furthermore, we assume
that the initial event was so large as to span the whole system,
i.e. $\rho_0(x)=1;\; 0\leq x<1$. 

Under the above assumptions there is basically one integral that
appears in Eq.\ (\ref{At-recursion}), which is
\begin{align}\label{I_nDef}
  I_n&:=\int\limits_0^1T^n(x)dx\notag\\
  &=\int\limits_0^1 (1-f)^n\Big(1-e^{-x/\sigma}\Big)^n dx\,,
\end{align}
and Eq.\ (\ref{At-recursion}) becomes
\begin{equation}
  A_t=1-I_t+\sum\limits_{k=1}^{t-1}I_{t-k}A_k\,.
\end{equation}
With the aid of the binomial expansion
$(1+a)^n=\sum_{k=0}^n\binom{n}{k}a^{k}$ we find
\begin{align}\label{I_nComp}
  I_n&=(1-f)^n\int\limits_0^1\sum\limits_{k=0}^n\binom{n}{k} 
          \Big(-e^{-x/\sigma}\Big)^k \notag\\
  &= (1-f)^n\bigg(1+ \sum\limits_{k=1}^n\binom{n}{k}(-1)^k\Big(
          \frac{\sigma}{k}-\frac{\sigma e^{-1/\sigma}}{k}\Big)\bigg)\notag\\
  &= \sigma(1-f)^n\bigg(1+ \sum\limits_{k=1}^n\binom{n}{k}\frac{(-1)^k}{k}\Big(
          1-e^{-1/\sigma}\Big)\bigg)\,.
\end{align}

We are now in the position to calculate the probability that an event
of size $s\geq s_1$ occurs at time $t$ after the initial big event. The
minimal stress value $\eta_{\min}$ that suffices on average to generate
such an event is the solution to the 
equation
\begin{equation}\label{eta_minEq}
  \int\limits_0^{\eta_{\min}}\!dx\,\rho_t(x)=s_1\,.
\end{equation}
The corresponding probability is $P_t(s\geq
s_1)=\exp(-\eta_{\min}/\sigma)$. We invert this expression and insert it
into Eq.~(\ref{eta_minEq}). The resulting equation determines
the probability $P_t(s\geq s_1)$:
\begin{equation}\label{P-Eq}
  \int\limits_0^{-\sigma\ln P_t(s\geq s_1)}\!dx\,\rho_t(x)=s_1\,.
\end{equation}
The integrals that appear after inserting $\rho_t(x)$ into the above
equation are very similar to the integral
$I_n$ defined in Eq.~(\ref{I_nDef}). We define
\begin{equation}
  J_n(P):=\int\limits_0^{-\sigma\ln P}T^n(x)\,dx\,.
\end{equation}
This integral can be taken in the same fashion as the calculation of $I_n$ in
  Eq.~(\ref{I_nComp}). We find
\begin{equation}
  J_n(P)=\sigma(1-f)^n\Big(-\ln P+\sum\limits_{k=1}^n\binom{n}{k} 
        \frac{(-1)^k}{k}(1-P)\Big)\,.
\end{equation}
Eq.~(\ref{P-Eq}) now becomes 
\begin{equation}\label{P-Eq2}
  J_t\Big(P_t(s\geq s_1)\Big)+\sum\limits_{k=1}^t J_{t-k}\Big(P_t(s\geq
  s_1)\Big)A_k=s_1\,.
\end{equation}
All the quantities which appear in this equation are given above in
analytic form. Therefore solving Eq.~(\ref{P-Eq2}) is simply a problem
of root-finding. With a computer-algebra program such as Mathematica,
the recursion relation for the constants $A_k$ as 
well as the sums that appear in the quantities $I_n$ and $J_n(P)$ can
be evaluated analytically if we restrict ourselves to moderate $t$.
Then the  only numerical calculation involved in the computation of
$P_t(s\geq s_1)$ is the calculation of the root of Eq.~(\ref {P-Eq2}).   
\end{appendix}

\newpage

\end{document}